\documentclass[twocolumn]{article} 

\usepackage{amssymb}
\usepackage{url}
\usepackage{hyperref}
\usepackage{authblk}
\usepackage{graphicx}

\title{Investigating solid $\alpha-^{15}$N$_{2}$ as a new source of ultra-cold neutrons}

\date{11 June 2013}

\author[1]{D.J. Salvat}
\author[1]{C.-Y. Liu}
\author[2]{E. Gutsmiedl}
\author[2]{S. Paul}
\author[3]{P. Geltenbort}
\author[3,4]{A. Orecchini}
\author[3]{H. Schober}

\affil[1]{Physics Department and Center for Exploration of Energy and Matter, Indiana University, Bloomington, Indiana 47405, USA}
\affil[2]{Physik-Department, Technische Universit\"{a}t Munchen, James-Franck-Str., D-85747 Garching, Germany}
\affil[3]{Institut Laue Langevin, 6 Rue Jules Horowitz, F-38042 Grenoble CEDEX9, France}
\affil[4]{Dipartimento di Fisica, Universit\`{a} di Perugia -- I-06123 Perugia, Italy}

\begin{document}

\maketitle

\begin{abstract}
        The dynamical structure factor of solid $^{15}$N$_{2}$ in the $\alpha$ phase ($T<35$K) is measured at the IN4 time-of-flight spectrometer
        at the Institut Laue Langevin, and the potential performance of this substance as a UCN converter is assessed.
        The cross-section to down-scatter neutrons to ultra-cold neutron energies is determined as a function of incident energy,
        as well as the up-scattering mean free path. The UCN production cross-section is found to be approximately $20\%$ of that of deuterium.
        However, UCN with energy $181$ neV have an up-scattering mean free path of $46$ cm at $T=5.9$ K, which is $\sim20$ times larger than deuterium.
        Therefore, a large volume $\alpha-^{15}$N$_{2}$ source may produce an improved UCN density if sufficient isotopic purity can be achieved.
\end{abstract}

\section{Introduction}

Ultra-Cold Neutrons (UCN) are neutrons with energies ($E \sim 100$ neV) low enough to be trapped by suitable material and magnetic bottles\cite{golub,ignatovich}.
Due to long storage times (several hundred seconds or more), precise measurements of the neutron's properties use UCN, and these measurements have implications for the theory of weak interactions
and physics beyond the standard model\cite{scalartensor,abele,nico}. Many of these experiments are statistically limited, and research of improved materials and technologies is
necessary to develop new and intense UCN sources.

Several planned UCN sources use solid deuterium (D$_{2}$) to down-scatter cold or thermal neutrons to UCN energies\cite{lanl,frm2,psi,pulstar}.
Since it was first proposed and investigated\cite{film,pulsed,serebrov}, considerable recent effort has gone into understanding UCN production in D$_{2}$\cite{freid2,liud2,atchisond2,atchisond2prod}.
Deuterium is chosen for its low neutron absorption cross-section,
low incoherent scattering cross-section (to minimize UCN elastic scattering within the source), and the presence of numerous phonon modes which can inelastically scatter neutrons down to UCN energies.
However, aside from these phonon modes, the free rotation of D$_{2}$ molecules can cause incoherent up-scattering (i.e. loss) of UCN in the deuterium.
In order to suppress this up-scattering, the D$_{2}$ sample is converted to the $J=0$ (ortho) state using paramagnetic catalysts\cite{freid2,liuortho}.

Solid $\alpha-^{15}$N$_{2}$ is a potential alternative to deuterium: its absorption cross-section is only $5$\% of that of D$_{2}$, and it has a negligible incoherent
scattering cross-section. Additionally, rotation of the N$_{2}$ molecules in the lattice is inhibited due to the anisotropy of the N$_{2}$ inter-molecular potential.
This leads to dispersive modes for the rotational degrees of freedom (librons) which provide additional channels for neutron down-scattering,
and eliminates the rotational incoherent up-scattering prevalent in D$_{2}$.

Due to the primarily classical nature of its inter-molecular potential, the anharmonicity of the angular degrees of freedom, and phonon/libron coupling,
solid nitrogen has been a testing-ground for \textit{ab initio} lattice and molecular dynamics calculations\cite{n2review}.
Several models exist which utilize inter-molecular or atom-atom potentials of various forms\cite{lattice1,lattice2,md1,md2,libration,anisotropy},
and anharmonic effects have also been investigated experimentally and theoretically\cite{anharmonicity1,anharmonicity2,anharmonicity3,anharmonicity4}.
Past interest notwithstanding, there is little data on the dynamical structure factor of $\alpha$-N$_{2}$ from neutron scattering\cite{kjems}.

In this letter, we present a measurement of the dynamics of $\alpha-^{15}$N$_{2}$ using neutron scattering in order to investigate the performance of nitrogen as a UCN converter.
The dynamical structure factor $S(q,E)$ is measured using the time-of-flight (TOF) technique;
the total scattering cross-section is then analyzed to determine the absolute intensity of the scattering. From this, the cross-section to convert
neutrons to UCN energy is computed.

\section{Experiment}

The dynamical structure factor is measured using the IN4 TOF spectrometer at the Institut Laue Langevin (ILL) using incident neutrons of wavelength $\lambda=2.2 \AA$\cite{in4}.
The sample environment, cryostat, and sample control system are the same as that used for previous studies of oxygen and deuterium as UCN converters\cite{freid2,gutsmiedlo2}.
This provides a means of direct comparison between the different converters.
Gaseous $^{15}$N$_{2}$ (99.4\% isotopic purity) is introduced into the $2$ mm annular cell through a needle valve, and condensed using the ILL's orange helium-exchange cryostat
(see \cite{freid2,gutsmiedlo2} for details). The detector array is calibrated using an isotropically scattering vanadium sample.
The dynamical structure factor and related quantities are computed using the ILL's standard data treatment software LAMP\cite{lamp}.

Total scattering from the sample cell is measured while condensing nitrogen to assure that the cell is completely filled,
and the sample temperature is then lowered to the solid $\alpha$-phase. In order to ensure that the nitrogen sample did not consist of macro-crystals,
the elastic scattering intensity is analyzed for various annealing times at the $\alpha-\beta$ phase transition. We find little variation in the elastic peaks for different annealing times.

Data are then acquired with nitrogen at temperatures of $5.9$, $11$, $15$, $20$, and $25$ K.
A background measurement of the empty cell is performed at $11$ K. The inelastic scattering background due to the aluminum phonons is temperature dependent, especially on the anti-Stokes ($E<0$) side,
according to the Bose-Einstein occupation factor $n_{BE}(T,E)=\left[ \exp \left( E / kT \right) - 1 \right]^{-1}$. 
We thus rescale the background intensity by the relative change in occupation number (as a function of $E$) prior to subtracting it from the nitrogen data.
This is particularly important for computing UCN up-scattering mean free path for $T<11$ K (see below), where it is a $\sim20$\% effect if uncorrected.

We also perform measurements using $^{14}$N$_{2}$, and find that the dynamical structure is the same as that for $^{15}$N$_{2}$, with a difference in scattering intensity commensurate with
the isotopic difference in total scattering cross-section ($\sigma_{scat 14} / \sigma_{scat 15} \approx 2.2$). For this reason, data with $^{14}$N$_{2}$ are used for the determination
of the temperature dependence of the dynamical structure factor due to the accumulation of better statistics. The results are scaled by the above factor to determine the absolute $^{15}$N$_{2}$ cross-sections.

\section{Results}

\begin{figure}[h]
        \includegraphics[width=1.0\linewidth]{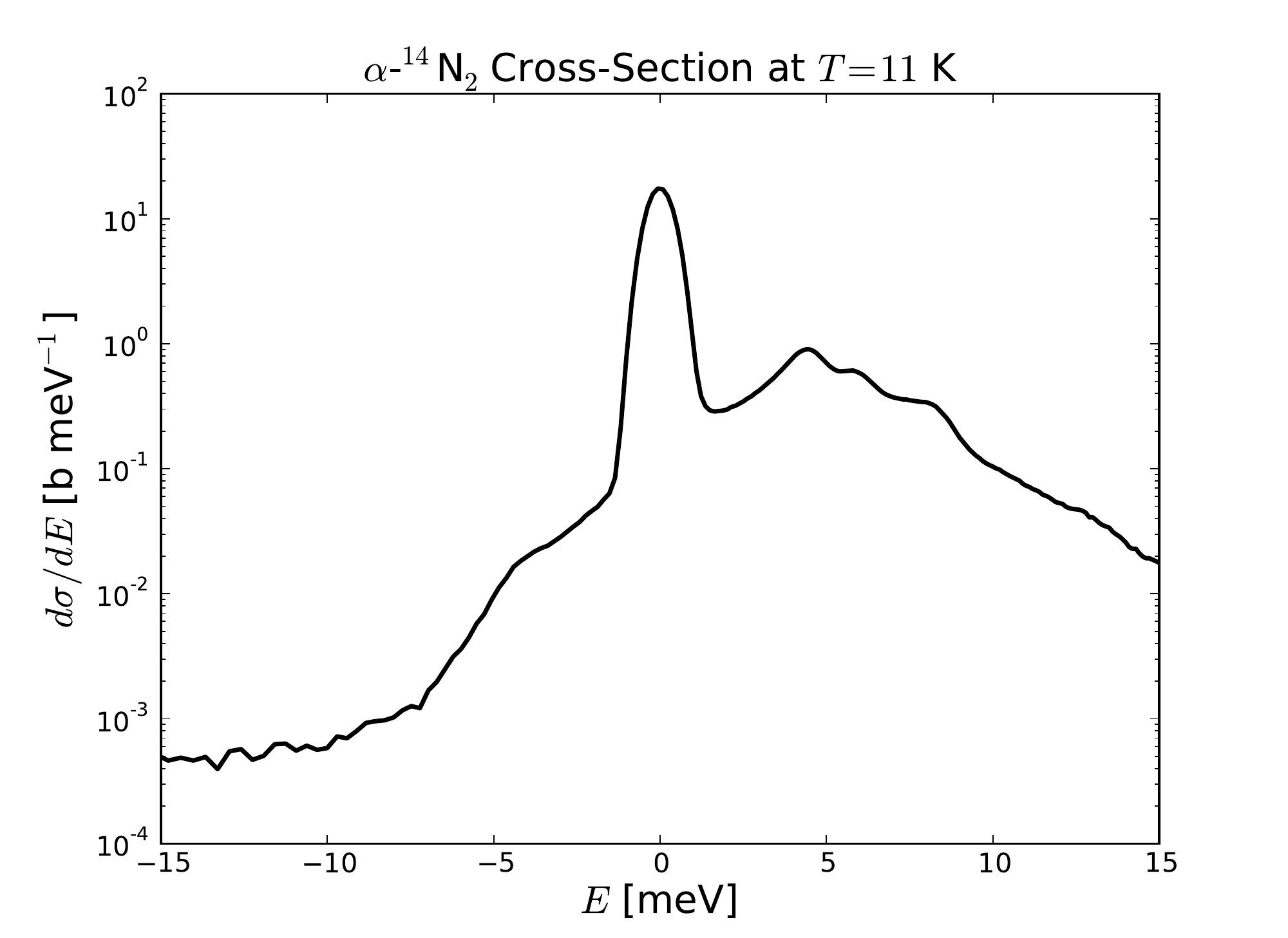}
        \caption{The differential scattering cross-section versus $E$. $E>0$ corresponds to energy loss, and $E<0$ to energy gain. The vertical scale is set by integrating
        to the total scattering cross-section (see text).}
        \label{dsde}
\end{figure}

The differential scattering cross-section $d \sigma / d E$ is shown in fig. \ref{dsde}. The vertical scale factor $\kappa$ is set by demanding that
\begin{equation}
        \kappa \int \left( \frac{d \sigma}{d E} \right)_{data} d E = \sigma_{scat}
\end{equation}
where $\sigma_{scat}$ is the total scattering cross-section per molecule.
We find $\kappa = 36.6 \pm 7.3$ b, with the $20$\% uncertainty due primarily to the unknown contribution of scattering intensity outside the kinematic range of the instrument.
The factor $\kappa$ has units of barns so as to convert the $d\sigma/dE$ data (which has an arbitrary normalization) into a cross-section.
This normalization factor is used below to calculate the absolute UCN production and up-scattering cross-sections.

\begin{figure}[h]
        \includegraphics[width=1.0\linewidth]{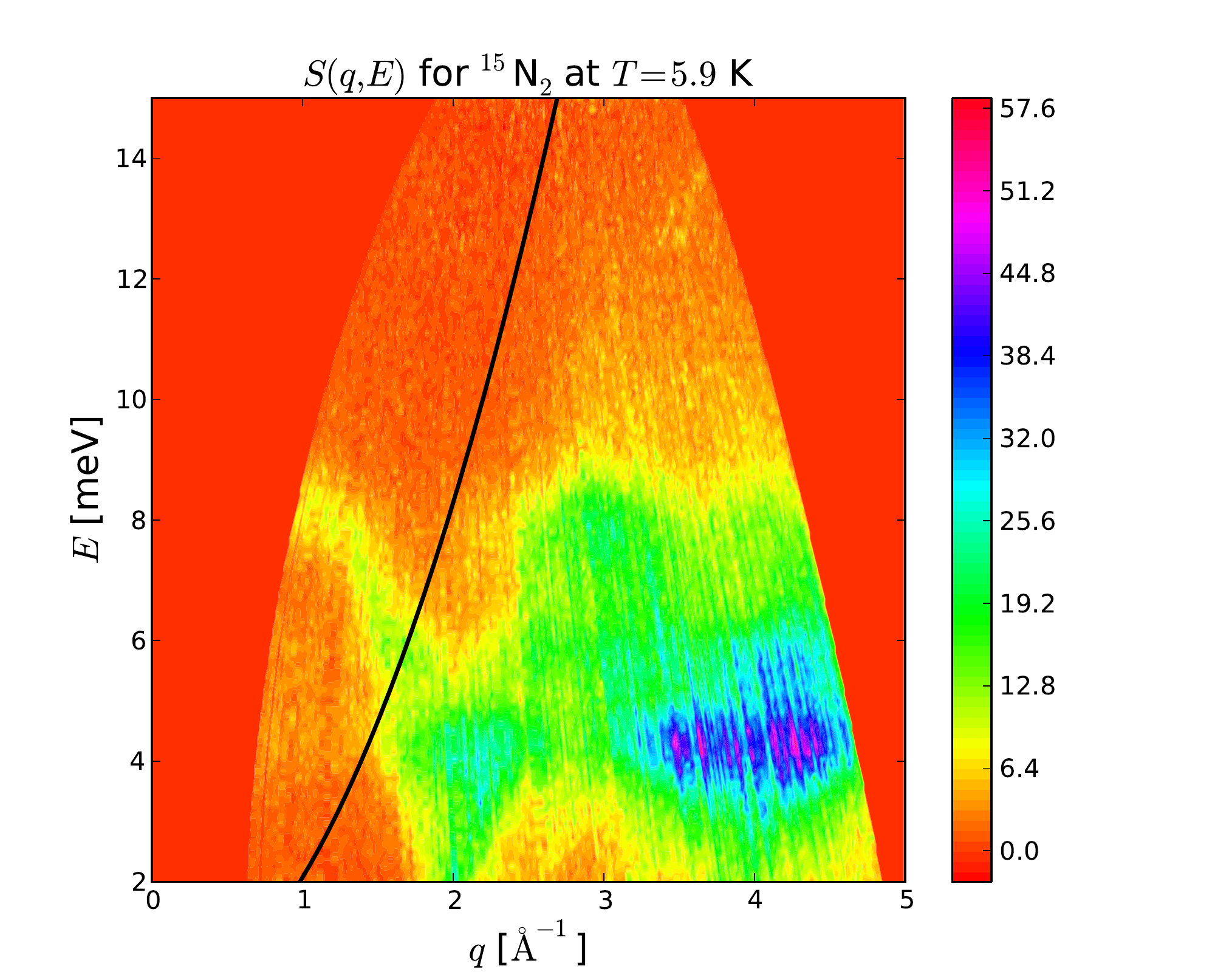}
        \caption{The dynamical structure factor of polycrystalline $\alpha-^{15}$N$_{2}$. The color scale is set by integrating the measured differential cross-section
        and equating it to the total scattering cross-section. The black line corresponds to the UCN production curve given by eqn \ref{parabola}.}
        \label{sqe}
\end{figure}

The dynamical structure factor $S(q,E)$ (averaged over orientations of $\vec{q}$ due to the polycrystallinity of the sample) is shown in fig. \ref{sqe}.
The intensity at $q=1 \AA^{-1}$ has peaks at $E=4.5$, $5.7$, and $8$ meV, in reasonable agreement with the T$_{g}$, A$_{u}$, and T$_{u}$ modes predicted by Kjems and Dolling\cite{kjems}.
However, it is difficult to comprehensively compare the data to various dynamical models due to the low energy resolution of the TOF method, as well as the polycrystalline
nature of the sample.

\begin{figure}[h]
        \includegraphics[width=1.0\linewidth]{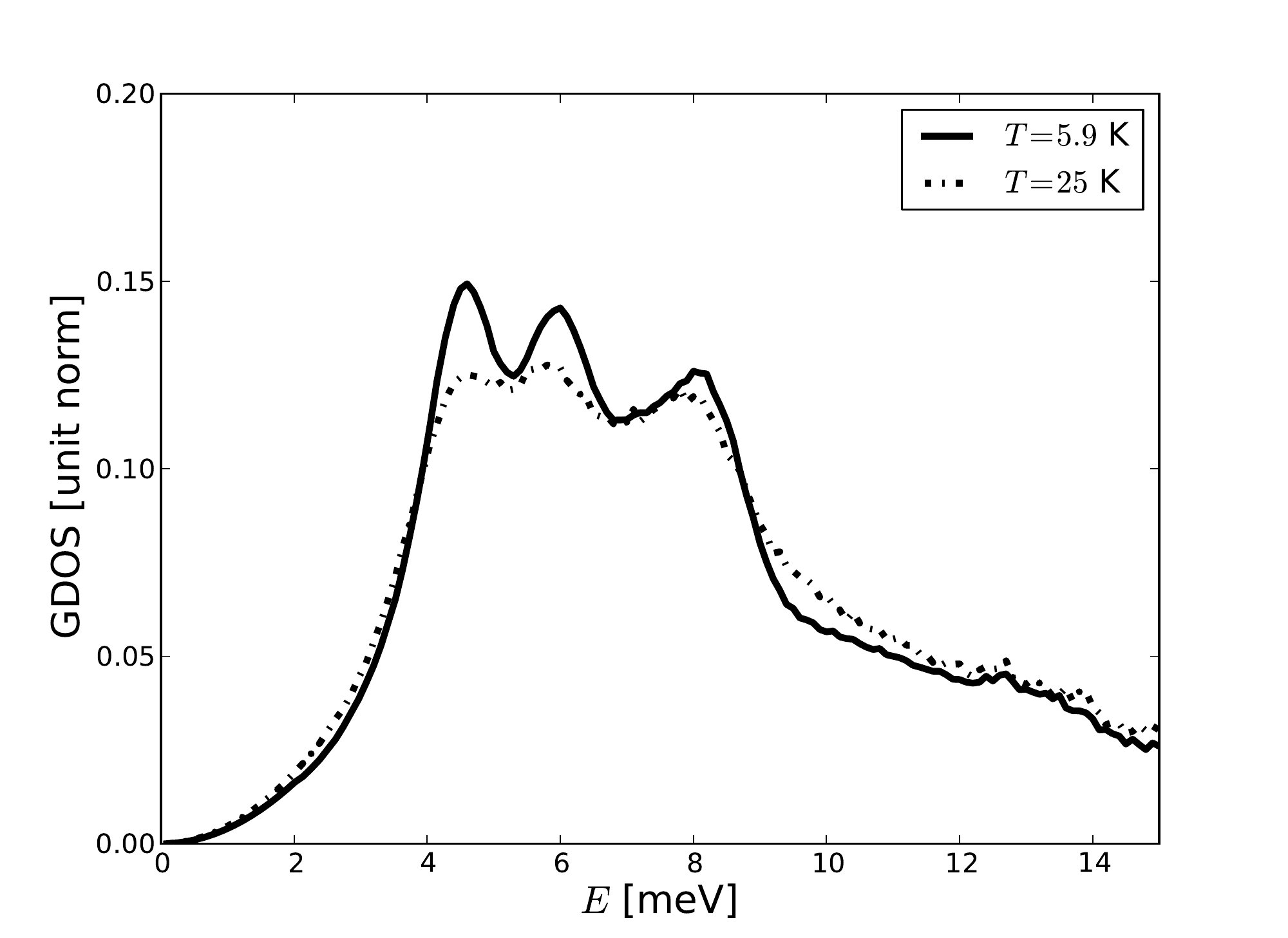}
        \caption{The GDOS for solid nitrogen at two temperatures in the $\alpha$-phase.
        Peak broadening at higher temperature is observed.}
        \label{gdos}
\end{figure}

Fig. \ref{gdos} shows the Generalized Density-Of-States (GDOS), which exhibits prominent peaks from one-phonon states,
and broad low-lying intensity due to multi-phonon processes. Elastic contamination is removed by fitting the GDOS to $E^{2}$ for $E\lesssim2$ meV.
The peaks in the $4$ to $8$ meV range broaden with increasing temperature due to phonon and libron conversion
and dephasing, which has been studied experimentally in detail using infrared and Raman spectroscopy\cite{anharmonicity2,anharmonicity3}.

\section{Discussion}\label{discussion}

The black line in fig. \ref{sqe} corresponds to the kinematical condition necessary for UCN down-scattering.
As UCN energies are much less than cold and thermal neutron energies, we take the UCN energy $E_{UCN} \approx 0$
and arrive at the following relation for the wave-vector transfer $q = \left| \vec{k}_{i} - \vec{k}_{f} \right|$ and energy transfer $E = E_{i} - E_{f}$:
\begin{equation}
        E = \pm \frac{\hbar^{2} q^2}{2 m_{n}}
        \label{parabola}
\end{equation}
with the plus sign for UCN production, and minus sign for UCN up-scattering.

The intensity of $S(q,E)$ parameterized along the curve given in eqn. \ref{parabola} is then proportional to the cross-section to nearly ``stop'' a cold neutron, thereby converting it to a UCN.
The UCN production cross-section versus incident energy integrated over final UCN energy $E_{f}$ from $0$ to some maximum UCN energy $E_{max}$ is given by

\begin{equation}
        \sigma_{UCN}(E_{i}) = \frac{2}{3} \kappa E_{max}^{3/2} \cdot \frac{1}{\sqrt{E_{i}}} S \left( \sqrt{2 m_{n} E_{i}}/\hbar,E_{i} \right).
\end{equation}

Upon escaping the nitrogen volume, UCN are boosted by the solid $^{15}$N$_{2}$ Fermi potential of $\sim69$ neV. We choose $E_{f}=181$ neV, which corresponds to a trappable energy
of $250$ neV, commensurate with UCN production cross-section calculations in \cite{gutsmiedlo2}.

\begin{figure}[h]
        \includegraphics[width=1.0\linewidth]{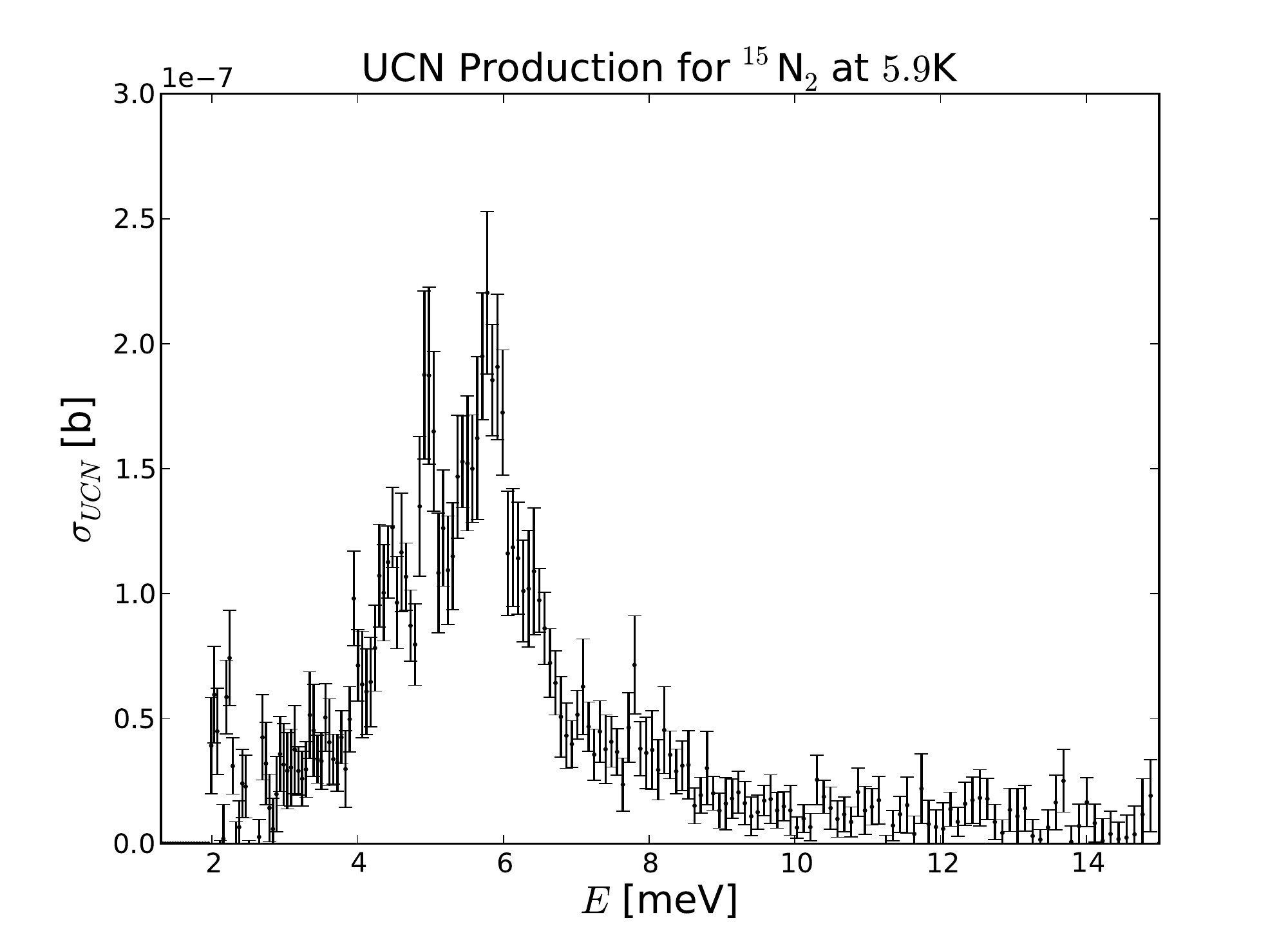}
        \caption{The UCN production cross-section in $^{15}$N$_{2} $for $E_{UCN} \leq 181$ neV versus incident neutron energy.
        The inelastic cross-section for $E<2$ meV is difficult to determine due to elastic contamination, though is likely small compared to that of the energy range shown here.}
        \label{sigmaucn}
\end{figure}

The energy-dependent production cross-section is shown in fig. \ref{sigmaucn}. Production peaks near $6$ meV, most probably due to a combination of vibrational and librational
down-scattering channels. The production cross-section can vary with temperature due to the temperature dependence of normal mode frequencies and their widths in the $\alpha$-phase:
we find the variation in the cross-section to be no more than $18$\% in the range of $5$ to $25$ K (increasing slightly with increasing temperature).

\begin{figure}[h]
        \includegraphics[width=1.0\linewidth]{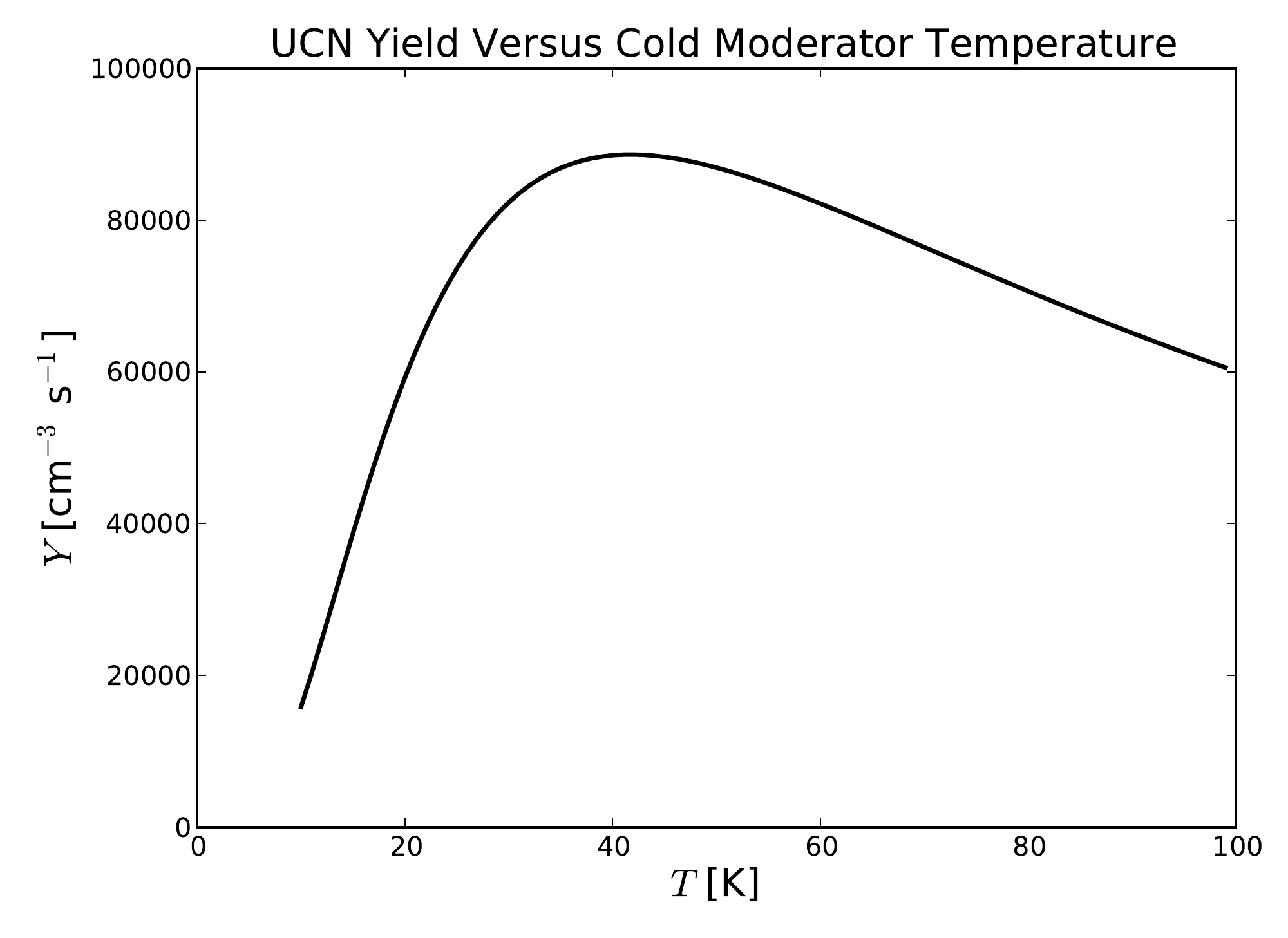}
        \caption{The UCN yield of $^{15}$N$_{2}$ for a neutron flux of $10^{14}$ cm$^{-2}$ s$^{-1}$ from a cold moderator at temperature $T$.
        The yield in UCN is optimized for incident neutrons at a temperature $T=40$ K.}
        \label{yield}
\end{figure}

From the above, an estimated rate of UCN production in a source using $^{15}$N$_{2}$ can be estimated, making reasonable assumptions about the incident neutron flux.
For a UCN source illuminated with the neutron flux from a cold moderator at temperature $T$, we define the UCN yield to be
\begin{equation}
        Y(T) = n \int{\phi \left(E,T\right) \sigma_{UCN}(E)dE}
\end{equation}
where $n = 2 \times 10^{-22}$ cm$^{-3}$ is the number density of $^{15}$N$_{2}$, and the cold neutron flux is given by
\begin{equation}
        \phi(E,T) = \phi_{0} \sqrt{\frac{4E}{\pi k^{3} T^{3}}} \exp \left( -E/kT\right).
\end{equation}
From this, we find an optimal moderator temperature of $40$ K (see fig. \ref{yield}). For a UCN source with total incident cold neutron flux flux $\phi_{0} = 10^{14}$ cm$^{-2}$ s$^{-1}$
at $T=40$ K, the yield is approximately $8.9\times 10^{4}$ UCN per second per cm$^{3}$ of source volume.

In order to extract a UCN from the source for use in an experiment, its mean free path $\lambda$ must be comparable to or larger than the thickness
of the source. The mean free path contains contributions from different processes:

\begin{equation}
        \lambda^{-1} = \lambda^{-1}_{up} + \lambda^{-1}_{abs} + \lambda^{-1}_{inc. el.}
\end{equation}
where $\lambda_{up}$ corresponds to UCN up-scattering to non-UCN energies, $\lambda_{abs}$ to nuclear absorption, and $\lambda_{inc. el.}$ to UCN elastically scattering
within the nitrogen volume. Even for relatively poor isotopic purity, the incoherent scattering mean free path is negligibly long.
The absorption mean free path is highly dependent on the isotopic purity, and purity $\sim 99.9$\% or higher is necessary for up-scattering to be the dominant contribution to the mean free path.

\begin{figure}[h]
        \includegraphics[width=1.0\linewidth]{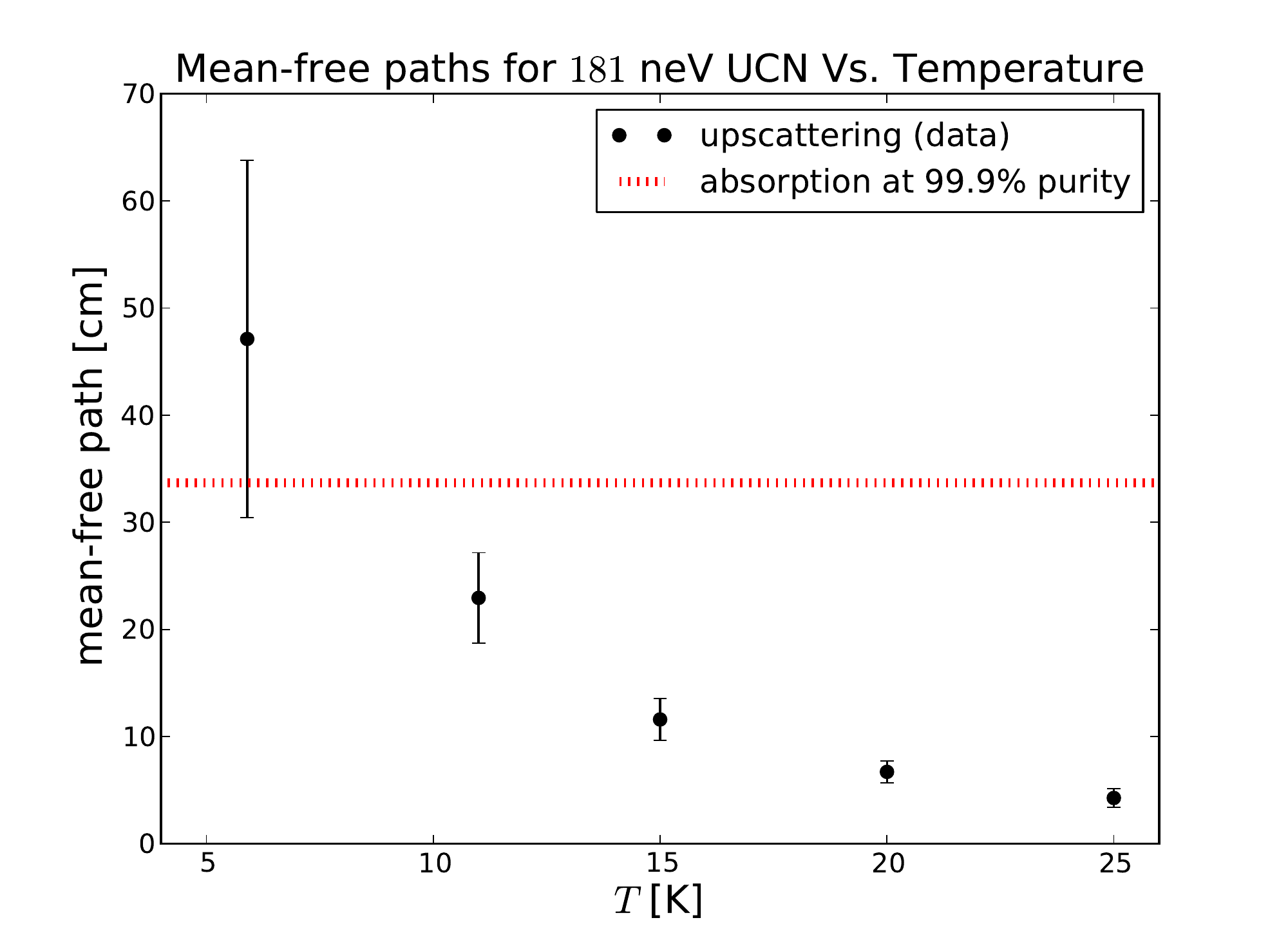}
        \caption{The mean free path $\lambda_{up}$ for UCN to up-scatter to non-UCN energies within the solid nitrogen volume. The error bars are statistical.}
        \label{mfp}
\end{figure}

The (temperature dependent) up-scattering cross-section of UCN in the nitrogen can be computed from the data by choosing the minus sign in eqn. \ref{parabola}
and integrating over the final state energy:
\begin{equation}
        \sigma_{up} = \kappa E_{0}^{-1/2} \int_{0}^{\infty} \sqrt{E} S \left( \sqrt{2 m_{n} E}/\hbar , -E \right) d E.
\end{equation}
We compute $\lambda_{up}=1/\sigma_{up}n$ for UCN with $E_{0}=181$ neV for a range of temperatures (see fig. \ref{mfp}).
We observe an increased $\lambda_{up}$ with decreasing temperature due to the smaller population of lattice modes that can induce up-scattering.
At $T=5.9$ K, the data suggest an up-scattering mean free path of approximately $46$ cm,
which is greater than the thickness of D$_{2}$ sources, such as the Los Alamos source ($\sim5$ cm)
or the PSI source ($\sim 15$ cm)\cite{lanl,psi}.
The thickness of a UCN source using $^{15}$N$_{2}$ could therefore be made larger than a D$_{2}$ source, the latter of which is limited by its $8$ cm incoherent
scattering mean free path.

\section{Conclusions}

The dynamical structure factor of solid $\alpha-^{15}$N$_{2}$ has been measured over a range of temperatures. No appreciable difference in dynamics is observed
between $^{14}$N$_{2}$ and $^{15}$N$_{2}$, and there is reasonable qualitative agreement between our data and previous studies of the dynamics of solid $\alpha$-nitrogen.
However, a higher resolution neutron scattering technique would be necessary to thoroughly investigate the rich structure of the vibrational and librational modes.

The absolute UCN production cross-section is determined by normalizing the scattered intensity to the total molecular cross-section of the sample.
The cross-section peaks strongly near $6$ meV, and the optimal incident cold neutron temperature is $40$ K.
The measured cross-section is found to be somewhat lower than that of D$_{2}$ and O$_{2}$ (see \cite{gutsmiedlo2}). However,
we observe an up-scattering mean free path substantially longer than that of D$_{2}$. Thus, for sufficiently high isotopic purity,
an $\alpha-^{15}$N$_{2}$ source could be large while maintaining efficient UCN extraction to compensate for the lower production cross-section.

Further, a nitrogen-based source may benefit from operating temperatures below those used here, if the up-scattering cross-section can be further
reduced at lower temperatures ($\sim1$ K). While deuterium is ultimately limited by its temperature independent
absorption and incoherent scattering mean free paths, further improved isotopic purity of $^{15}$N may make an even lower temperature nitrogen volume 
a promising candidate for a new UCN source. This motivates future work in performing direct measurements of slow neutron mean free paths
and UCN production rates in $^{15}$N$_{2}$.

\section{acknowledgments}

This work is supported by the cluster of excellence ``Origin and Structure of the Universe'' Exc. 153 and the Maier-Leibnitz-Laboratorium (MLL)
of the Technische Universit\"{a}t M\"{u}nchen and Ludwig-Maximilians-Universit\"{a}t M\"{u}nchen.
Author DJS is supported by the DOE Office of Science Graduate Fellowship Program (DOE SCGF),
made possible in part by the American Recovery and Reinvestment Act of 2009, administered by ORISE-ORAU under contract no. DE-AC05-06OR23100.

We thank S. Rols for the development of some of the TOFHR LAMP routines used in the analysis and for the management of the IN4 instrument.
We also thank O. Meulien for assistance with the sample preparation and gas handling system.

\end{document}